\documentclass[ams, prb, twocolumn, amssymb, floatfix, longbibliography]{revtex4-1}
\usepackage{amsmath}
\usepackage{tabularx}
\usepackage{bm}
\usepackage{euscript}
\usepackage{graphicx}
\usepackage{color}
\usepackage{amsfonts}
\usepackage{exscale}
\usepackage{amsbsy}
\usepackage{subfigure}
\usepackage{textcomp}
\usepackage{comment}
\usepackage{slashed}
\usepackage{hyperref}

\newcommand{\nn}{\nonumber}

\newcommand{\beq}{\begin{equation}}
\newcommand{\eeq}{\end{equation}}
\newcommand{\be}{\begin{eqnarray}}
\newcommand{\ee}{\end{eqnarray}}

\begin{document}

\title{Anisotropic quantum Hall states in the presence of interactions with fourfold rotational symmetry
}
\author{Prashant Kumar$^{1}$ and R. N. Bhatt$^{2}$}
\affiliation{$^1$Department of Physics, Princeton University, Princeton NJ 08544, USA\\
		$^2$Department of Electrical Engineering, Princeton University, Princeton NJ 08544, USA
}
\date{\today}

\begin{abstract}
	We study the effects of anisotropic interactions in the quantum Hall effect in the presence of a fourfold discrete rotational ($C_4$) symmetry. Employing the density matrix renormalization group  technique on an infinite cylinder geometry (iDMRG), we calculate the anisotropy response of the Laughlin state at $\nu=1/3$ and the composite-Fermi liquid (CFL) state at $\nu=1/2$. We find that the anisotropy transferred from the interaction potential to the $\nu=1/3$ state is stronger when compared to the complementary case of an anisotropic band. Further, the strength of anisotropy reduces as the interaction is made shorter ranged. Quite surprisingly, at $\nu=1/2$, the deformation in the CF Fermi-surface changes sign as the interaction range is reduced. Our results imply that the short-distance and long-distance parts of the interaction potential have opposite effects on the CFL state in the presence of $C_4$-symmetric anisotropy.
\end{abstract}

\maketitle




\section{Introduction}
The quantum Hall effect (QHE) is one of the prime examples of topological phenomena in condensed matter physics where electron-electron interactions play a leading role. In the literature, rotational symmetry is often imposed as a presumably benign assumption even though the experimental systems often break it. However, not only the quantum Hall states survive when the rotational symmetry is broken, it was shown in Ref. \onlinecite{Haldane2011} that the assumption of rotational symmetry also hides the geometric degree of freedom in the quantum Hall states. Since then, a lot of attention has been paid to anisotropic quantum Hall systems.\cite{Qiu2012,Yang2012,Wang2012,Maciejko2013,Papic2013,You2014,Balram2016,Johri2016,Ciftja2017,Gromov2017,Gromov2017b,Ippoliti2017,Ippoliti2017b,Ippoliti2017c,Yang2017,Yang2017b,Zhu2017,Ippoliti2018,Lee2018,Liu2018,Zhu2018,Krishna2019,Bhatt2020, Bergholtz2008}

Generally speaking, the continuous rotational symmetry in quantum Hall systems can be broken, while preserving the translational symmetry, in two different ways: through an anisotropy in the band dispersion and/or an anisotropic dielectric tensor. For the case of a twofold discrete symmetry ($C_2$), the two kinds of anisotropies can be related by a straightforward scaling of the cartesian coordinate axes. However, for higher order rotational symmetries such as the fourfold symmetry ($C_4$), they are distinct. The case of $C_4$-symmetric band anisotropy has been studied previously in Refs. \onlinecite{Ippoliti2017b,Krishna2019}. In this paper, we look at the latter case of a $C_4$-symmetric anisotropic interaction in the presence of an isotropic band.

We study the quantum Hall effect for various types of anisotropic interactions using the density-matrix renormalization group numerical technique on an infinite cylinder geometry (iDMRG). Although only Coulomb, dipolar and short-range interactions are relevant to experiments, we study generic power-law interaction potentials of the form $V(r)= 1/r^\eta$ which can provide more insight into how different interactions affect the quantum Hall states. 

We introduce anisotropy in an isotropic interaction potential by making its equipotential curves non-circular. Focusing on the Laughlin state at filling fraction $\nu=1/3$ and the composite-Fermi liquid (CFL) state at $\nu=1/2$, we numerically compute the anisotropy in these QH states as the exponent $\eta$ is varied. 

The rest of the paper is organized as follows. In section \ref{sec:model}, we introduce the model and explain the methods that are used to measure the anisotropy of QH states. In section \ref{sec:1_3}, we present the results for the Laughlin state at $\nu=1/3$. Section \ref{sec:1_2} deals with the CFL state at $\nu=1/2$. We conclude in section \ref{sec:discussion} with a discussion of the results and future directions.

\section{Model and methods\label{sec:model}}
The two-dimensional electron gas in a perpendicular magnetic field $B$ is usually modeled as a system of fermions with a quadratic dispersion interacting via density-density interactions. To study the quantum Hall states at $\nu=1/3$ and $\nu=1/2$, we make a simplifying assumption that the interactions are weak compared to the cyclotron energy so that we can confine ourselves to the lowest Landau level (LLL). 

We work on an infinite cylinder geometry in our iDMRG simulations. Such a system has the advantage of providing a continuous momentum-space variable along the axis of the cylinder. We choose this infinite direction to be the $y$-axis and the $x$-axis corresponds to the compact direction of the cylinder with periodic boundary conditions. The cylindrical geometry is manifest in the Landau gauge and the Hamiltonian is given by:
\begin{align}
	H &= \frac{1}{2} \int d^2 r d^2r'\ V(\bm r-\bm r')\ \overline{: \rho(\bm r) \rho(\bm r'):}\nn\\
	&= \frac{1}{2} \sum_{n,m,k} V_{mk}\ c^\dagger_{n+m} c^\dagger_{n+k} c_{n+m+k} c_n\\
	V_{mk} &\equiv \frac{e^{-\kappa^2m^2/2}}{L_x} \int \frac{dq_y}{2\pi}\ V\left(\kappa m, q_y\right) e^{i\kappa k q_y - q_y^2/2}\label{eq:Vmk_definition}\\
	\rho (\bm r) &\equiv c^\dagger(\bm r) c(\bm r)
\end{align}
where $\kappa = 2\pi/L_x$,  $L_x$ represents the circumference of the cylinder, $n,m,k$ are integers labeling the Landau gauge orbitals with electron creation and annihilation operators $c^\dagger_m$ and $ c_m$ respectively, $V(q_x, q_y) = V(\bm q)$ is the density-density interaction potential in the momentum-space and the overline represents LLL projection. We have set the electric charge $e$, the reduced Planck's constant $\hbar$ and the magnetic length $l_B^2 \equiv 1/B$ equal to 1.

The natural way to break the rotational symmetry of Coulomb interactions is by making the dielectric tensor anisotropic. However, we would like to study more general interactions. Therefore, we'll introduce $C_4$-symmetric anisotropy in an otherwise isotropic interaction by making its equipotential curves non-circular. This is achieved by making the following substitution to a rotationally invariant potential $V_{\rm iso}(r)$: \footnote{This ansatz is different in form than the one used in Ref. \onlinecite{Ippoliti2017} in the case of a $C_4$-anisotropic band. However, to leading order in $\log \alpha$, the two are similar.}
\begin{align}
	 V(\bm r) &= V_{\rm iso}(r_{\rm ani} (r, \theta))\\
	r^2_{\rm ani} (r, \theta) &\equiv \frac{r^2}{2} \left[\left(\alpha + \frac{1}{\alpha}\right) + \left(\alpha - \frac{1}{\alpha}\right)\cos 4\theta \right] \label{eq:ani_prescription}
\end{align}
where $x = r\cos\theta,\ y= r\sin\theta$ and $\alpha$ quantifies the strength of anisotropy. The equipotentials of $V(\bm r)$ are given by $r_{\rm ani} (r, \theta) = \text{constant}$. The interaction potential is Fourier-transformed to momentum space and we obtain the LLL Hamiltonian using Eq. \eqref{eq:Vmk_definition}.

An interaction that breaks the two-dimensional rotational symmetry down to an $N$-fold discrete symmetry ($C_N$) introduces an anisotropy of the same order in the correlations of a quantum Hall ground state while not altering its topological content. This is true as long as the rotational symmetry is not further broken spontaneously and the energy spectrum remains gapped. It is also known that a (gapless) Fermi-liquid is not altered by anisotropy, we therefore assume that the same is true for the CFL state at $\nu=1/2$.

To measure the amount of anisotropy transferred to the quantum Hall ground states, we analyze the guiding-center static structure factor $S(\bm q)$ that can be readily calculated in our iDMRG simulations:
\begin{align}
	S(\bm q) &\equiv \frac{1}{N_e} \langle \delta\overline{ \rho} (\bm q)\delta\overline{ \rho} (-\bm q)\rangle\\
	\bar\rho(\bm q) &\equiv e^{i q_x q_y/2} \sum_{n} e^{i \kappa n q_y} c^\dagger_{n} c_{n+q_x/\kappa}
\end{align}
where $\bar\rho(\bm q)$ is the guiding-center density operator and $\delta\overline{ \rho} (\bm q) \equiv \overline{ \rho} (\bm q) - \langle \overline{ \rho} (\bm q)\rangle$. In a more familiar first quantized representation, $\bar\rho(\bm q)$ is given by:
\begin{align}
	\bar\rho(\bm q) = \sum_{j=1}^{N_e} e^{-i\bm q.\bm R_j}
\end{align}
where $N_e$ is the number of electrons and $\bm R_j$ is the guiding center operator of the $j^{th}$ electron.

$S(\bm q)$ encodes the anisotropy of the quantum Hall states in various ways. We'll use two different methods developed in Refs. \onlinecite{Ippoliti2017, Krishna2019} that utilize different regimes of the wavevector $\bm q$ to quantify the anisotropy: one via the small-$q$ behavior of $S(\bm q)$ and the other using certain features at nonzero-$q$.

\subsection{$q\rightarrow 0$ limit of $S(\mathbf{q})$}

For an isotropic and gapped quantum Hall liquid, in the limit $q \rightarrow 0$, the static structure factor behaves as $S(q) \approx c_0 q^4$\cite{Girvin1986}. With broken rotational symmetry, it generalizes to $S(\bm q) \approx c(\alpha,\phi) q^4$, where $\tan\phi = q_y/q_x$. In our quasi-1D iDMRG simulations, we can determine the small-$q$ behavior only along the $y$-direction. Altering the anisotropy from $\alpha$ to $1/\alpha$ amounts to a $\pi/4$ rotation (see Eq. \eqref{eq:ani_prescription}). We therefore use this pair $(\alpha, 1/\alpha)$ to obtain the anisotropy of the QH state. To this end, we make the following ansatz motivated by Eq. \eqref{eq:ani_prescription}:
\begin{align}
	S(\bm q) &\sim  S_{\rm iso}(q_{\rm ani}(q, \phi))\\
	q_{\rm ani}^2(q, \phi) &= \frac{q^2}{2 } \left[\left(\alpha_{\rm QH} + \dfrac{1}{\alpha_{\rm QH}}\right) + \left(\alpha_{\rm QH} - \dfrac{1}{\alpha_{\rm QH}}\right)\cos 4\phi \right]\label{eq:ani_QH_measure}\\
	c(\alpha, \phi) &= c_0(\alpha)\frac{q_{\rm ani}^4(q,\phi)}{q^4}
\end{align}
where $c_0(\alpha) = c_0(1/\alpha)$ and $\alpha_{\rm QH}$ corresponds to the anisotropy of the quantum Hall liquid. Also, $S_{\rm iso}(q)$ represents the static structure factor of the QH system with isotropic interaction $V_{\rm iso}(r)$ while $S(\bm q)$ corresponds to the anisotropic system obtained by the replacement prescribed in Eq. \eqref{eq:ani_prescription}. Using $\alpha_{\rm QH} \equiv \alpha_{\rm QH}(\alpha) = 1/\alpha_{\rm QH}(1/\alpha)$, the anisotropy can be calculated as follows:
\begin{align}
	\alpha_{\rm QH}(\alpha) &= \sqrt{\frac{\alpha_{\rm QH}(\alpha)}{\alpha_{\rm QH}(1/\alpha)}} = \left(\frac{\lambda(\alpha)}{\lambda(1/\alpha)}\right)^{1/4} \label{eq:ani_small_q}\\
	\lambda(\alpha) &\equiv c(\alpha, \pi/2)
\end{align}

\subsection{Features of $S(\mathbf{q})$ at $q>0$}

In addition to the small-$q$ behavior, the static structure factor has features at $q>0$ such as the singularities arising from particle-hole excitations in a composite-Fermi sea at $\nu=1/2$, or the maxima in the structure factor at $\nu=1/3$ corresponding to the magnetoroton minimum.\cite{Girvin1986} We assume that such features lie along the curves: $q^2_{\rm ani}(q, \phi) =$ constant in the 2D limit.\footnote{In the quasi-1D cylinder geometry used in our iDMRG simulations, there can be an additional $C_2$-rotationally invariant contribution that would vanish in the infinite circumference limit.} Therefore, after obtaining the $q$-values of these features for various values of $\phi$ or $\alpha$, we'll use the following equation to determine the anisotropy:
\begin{gather}
q^2(\phi) \propto \frac{2}{\left(\alpha_{\rm QH} + \dfrac{1}{\alpha_{\rm QH}}\right) + \left(\alpha_{\rm QH} - \dfrac{1}{\alpha_{\rm QH}}\right)\cos 4\phi}\label{eq:ani_finite_q}
\end{gather}

We note that the anisotropy in the interaction potential is defined in real-space as may be expected for an anisotropic dielectric tensor or interaction between multipoles.
On the other hand, the anisotropy of the QH system is measured in the momentum-space because it corresponds to the physical quantities that may be measured in an experiment, such as the shape of a Fermi-sea at $\nu=1/2$ or the magnetoroton minimum.

\section{$C_4$-symmetric interaction anisotropy at $\nu=1/3$\label{sec:1_3}}
In this section, we investigate the transfer of anisotropy from the interaction potential to the Laughlin state at $\nu=1/3$ for power-law interactions of the form:
\begin{gather}
	V(\bm r) = \frac{e^{-r^2/2\xi^2}}{r^\eta}\label{eq:pow_law_interaction}
\end{gather}
where we have introduced a Gaussian-envelope with a range $\xi$. We choose $\xi=6\ell_B$ in this paper which is big enough to study long-range interactions while small enough to avoid finite size effects that can arise from the finite circumference in the cylindrical geometry.\cite{Geraedtsetal2015} Also, for $\eta>1$, we introduce a small smoothing factor $\Delta = 0.001$ by replacing $r^\eta \rightarrow \left(r^2+\Delta^2\right)^{\eta/2}$ in the denominator. This has been found to have minor effects on the results.\cite{Krishna2019}

\begin{figure}
	\centering
	\includegraphics[width=3.3in]{"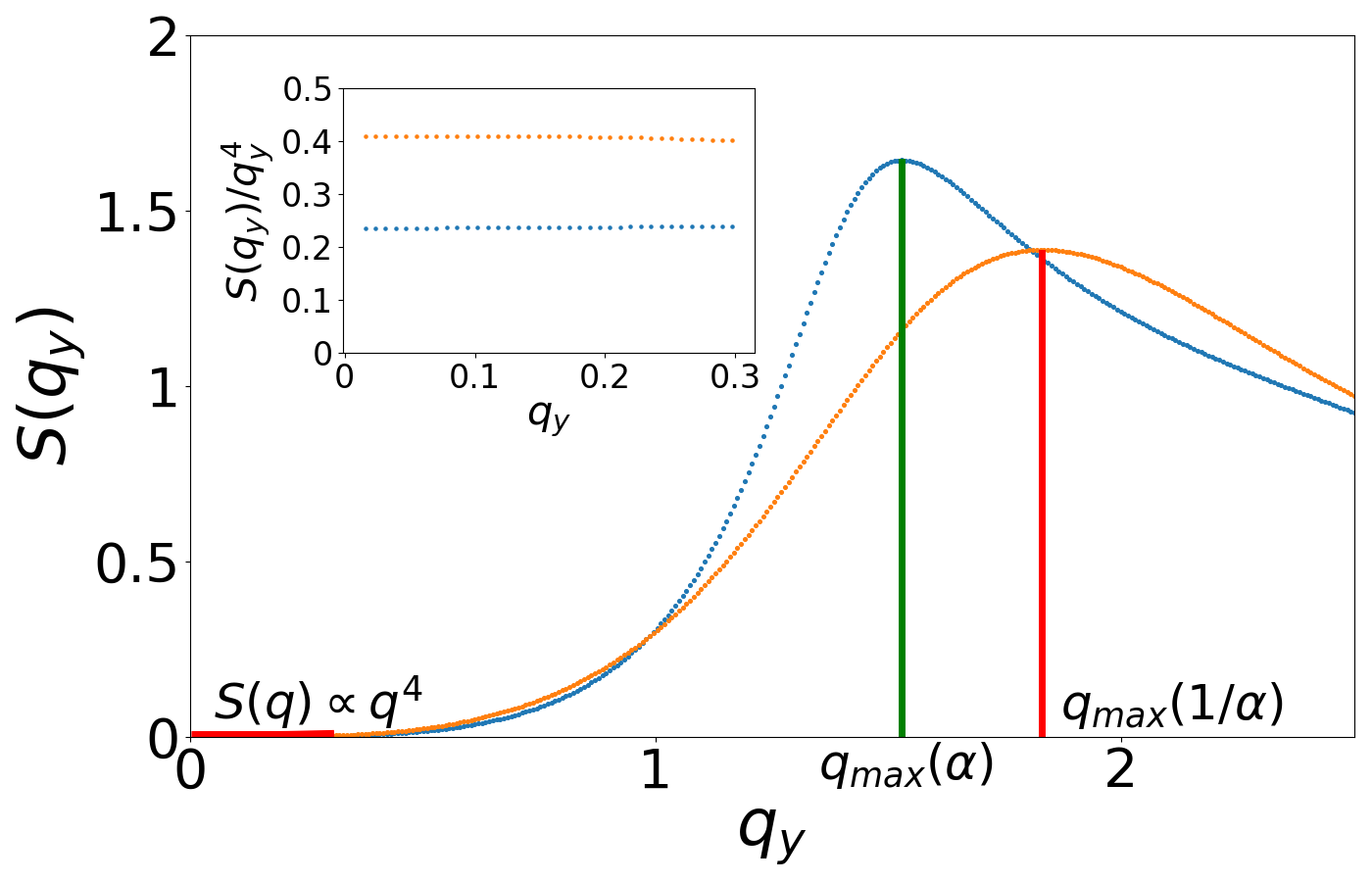"}
	\caption{The static structure factor $S(q_y) \equiv S(\bm q = (0, q_y))$ vs. $q_y$ of the Laughlin state at $\nu=1/3$ for $1/r$ interactions. The blue and the orange dotted-curves correspond to the $(\alpha$, $1/\alpha)$ pair of $C_4$-symmetric interaction-anisotropies with $\alpha=1.5$. The red curve at small-$q$ is a fit to the equation $S(q_y) = \lambda q_y^4$ with $\lambda(\alpha)=0.2359$ and $\lambda(1/\alpha) = 0.4108$. The inset shows $S(q_y)/q_y^4$ for the pair of curves converging to $\lambda(\alpha)$ and $\lambda(1/\alpha)$. Further, $q_{\rm max}$ corresponds to the location of the maxima in the structure factor at $q_{\rm max}(\alpha) = 1.53$ and $q_{\rm max}(1/\alpha) = 1.83$.}
	\label{fig:structure_factor_Laughlin}
\end{figure}

\begin{figure}
	\centering
	\includegraphics[width=2.5in]{"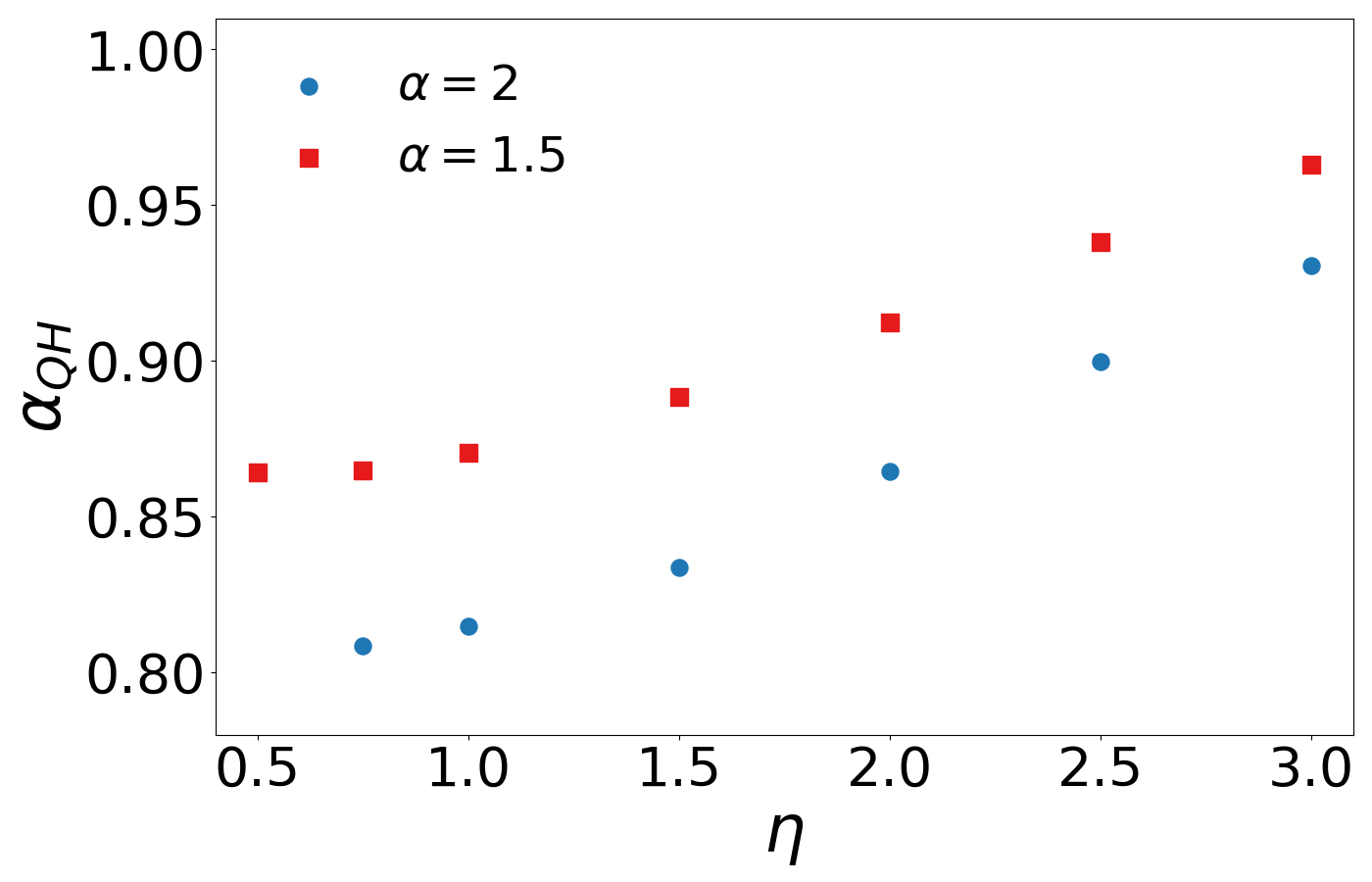"}
	
	(a)
	
	\includegraphics[width=2.5in]{"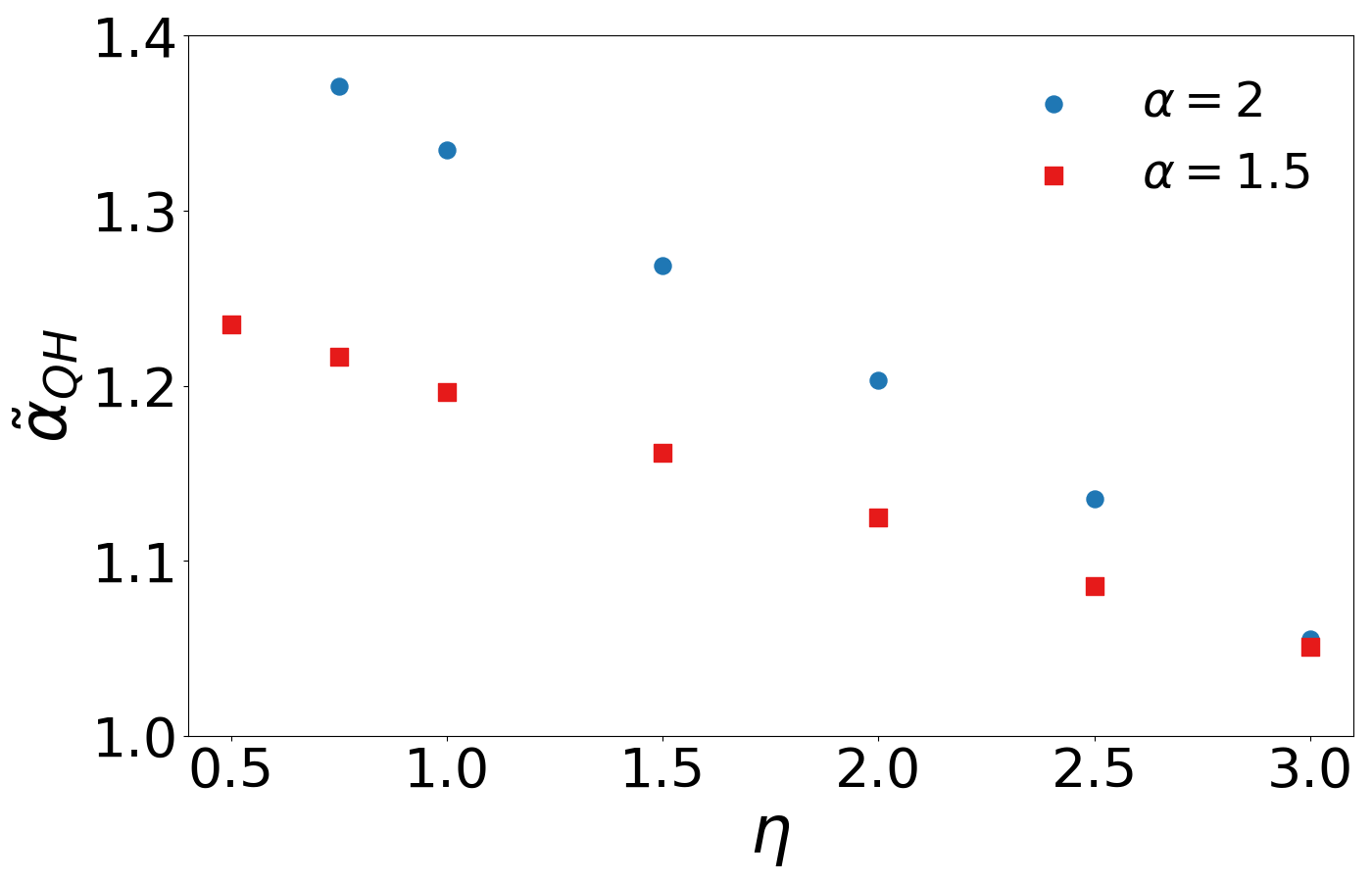"}
	
	(b)
	\caption{The dependence of anisotropy of the $\nu=1/3$ QH state on the power law exponent of the interaction $\eta$, i.e., $V(r) = 1/r^\eta$ using (a) $q \rightarrow 0$ behavior and (b) maxima of the static structure factor. In our iDMRG simulations, we used a cylinder of circumference $20\ell_B$ and bond dimension $\chi=8192$. The deformation in the QH state stays in the same orientation but reduces in magnitude as $\eta$ is increased. Also, the deformations in the small-q behavior and the magnetoroton minima are opposite in sign.}
	\label{fig:ani_QH}
\end{figure}

In Fig. \ref{fig:structure_factor_Laughlin}, we plot $S(\bm q)$ at $\nu=1/3$ at $\alpha=1.5$ and $q_x= 0$. At small-$q$, $S(q_x=0, q_y) \propto q_y^4$ and there is a maxima in the structure factor at $q_y = q_{\rm max}$ corresponding to the magnetoroton minimum. We determine the anisotropy in the small-$q$ regime $\alpha_{QH}$ using Eq. \eqref{eq:ani_small_q}. While for the finite-$q$ maxima, we make use of the following formula derived using Eq. \eqref{eq:ani_finite_q}:
\begin{gather}
	\tilde\alpha_{\rm QH}(\alpha) = \frac{q_{\rm max}(1/\alpha)}{q_{\rm max}(\alpha)}
\end{gather}

In Fig. \ref{fig:ani_QH}, we show the calculated anisotropy of the Laughlin state against the power law exponent $\eta$ using the two methods. The anisotropy is largest for small $\eta$ and reduces monotonically as $\eta$ is increased. Further, the deformation in the Laughlin state is always in the same direction independent of the exponent $\eta$.

\begin{figure}
	\centering
	\includegraphics[width=2.5in]{"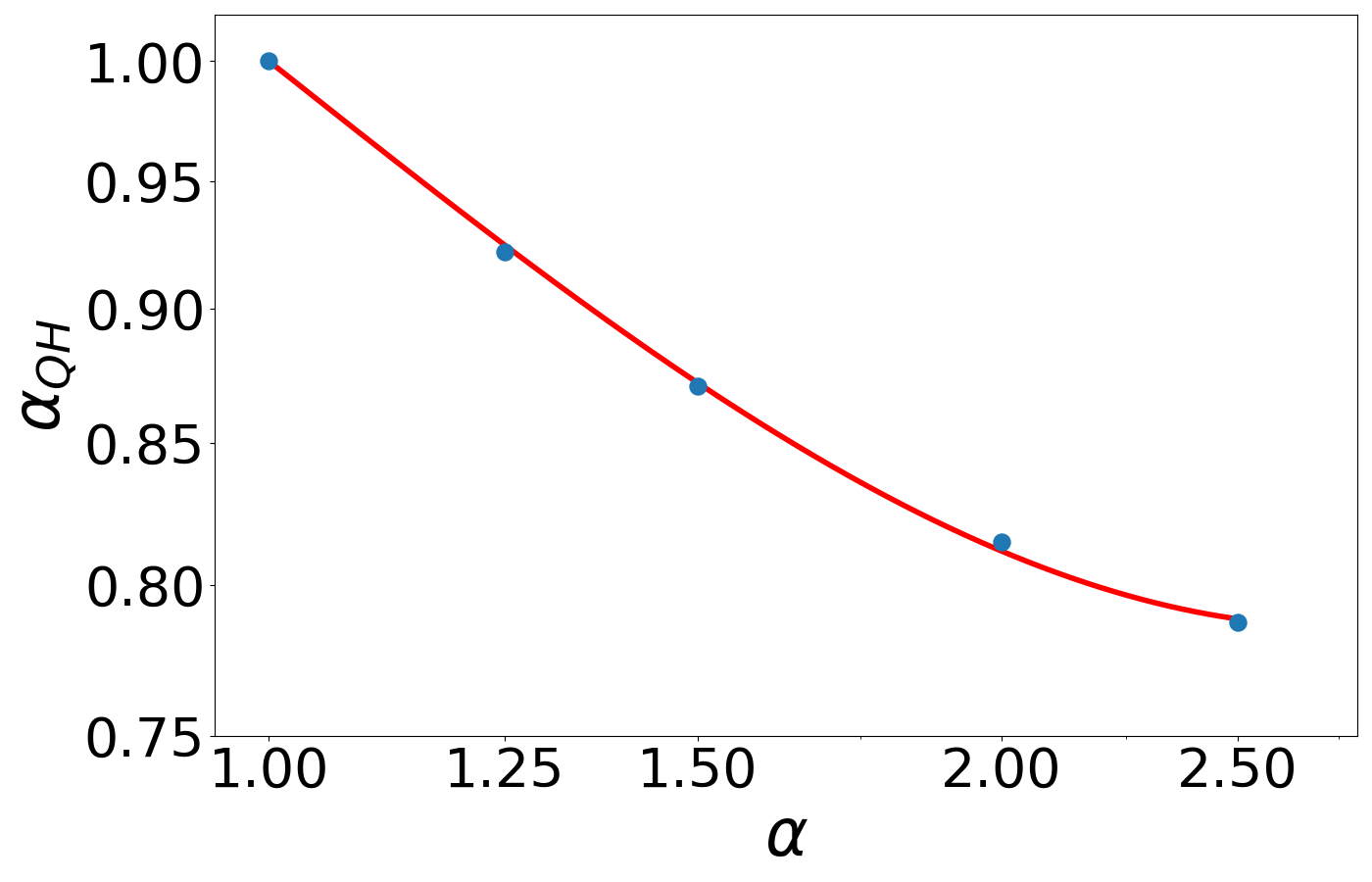"}
	
	(a)
	
	\includegraphics[width=2.5in]{"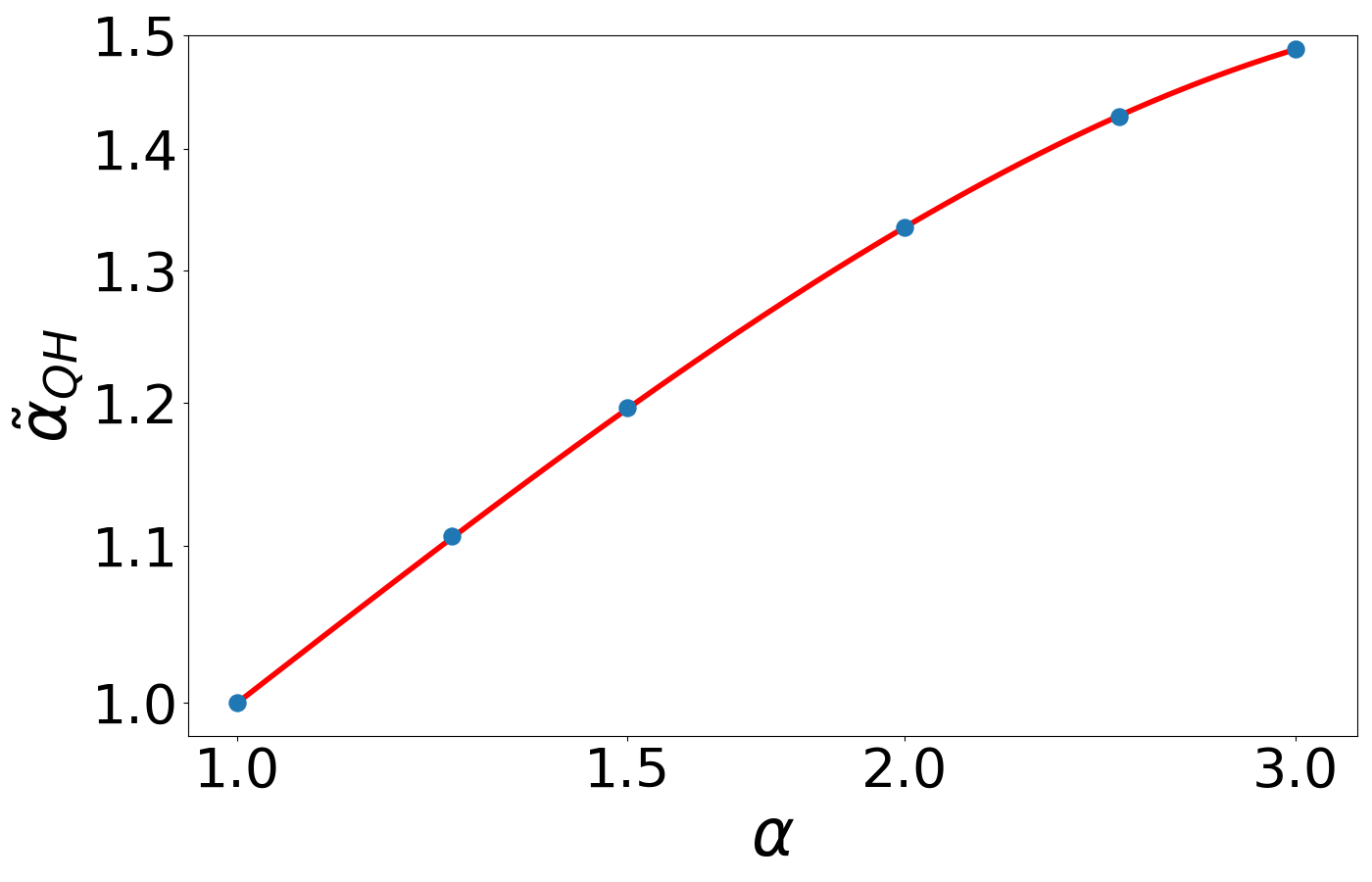"}
	
	(b)
	
	\includegraphics[width=2.5in]{"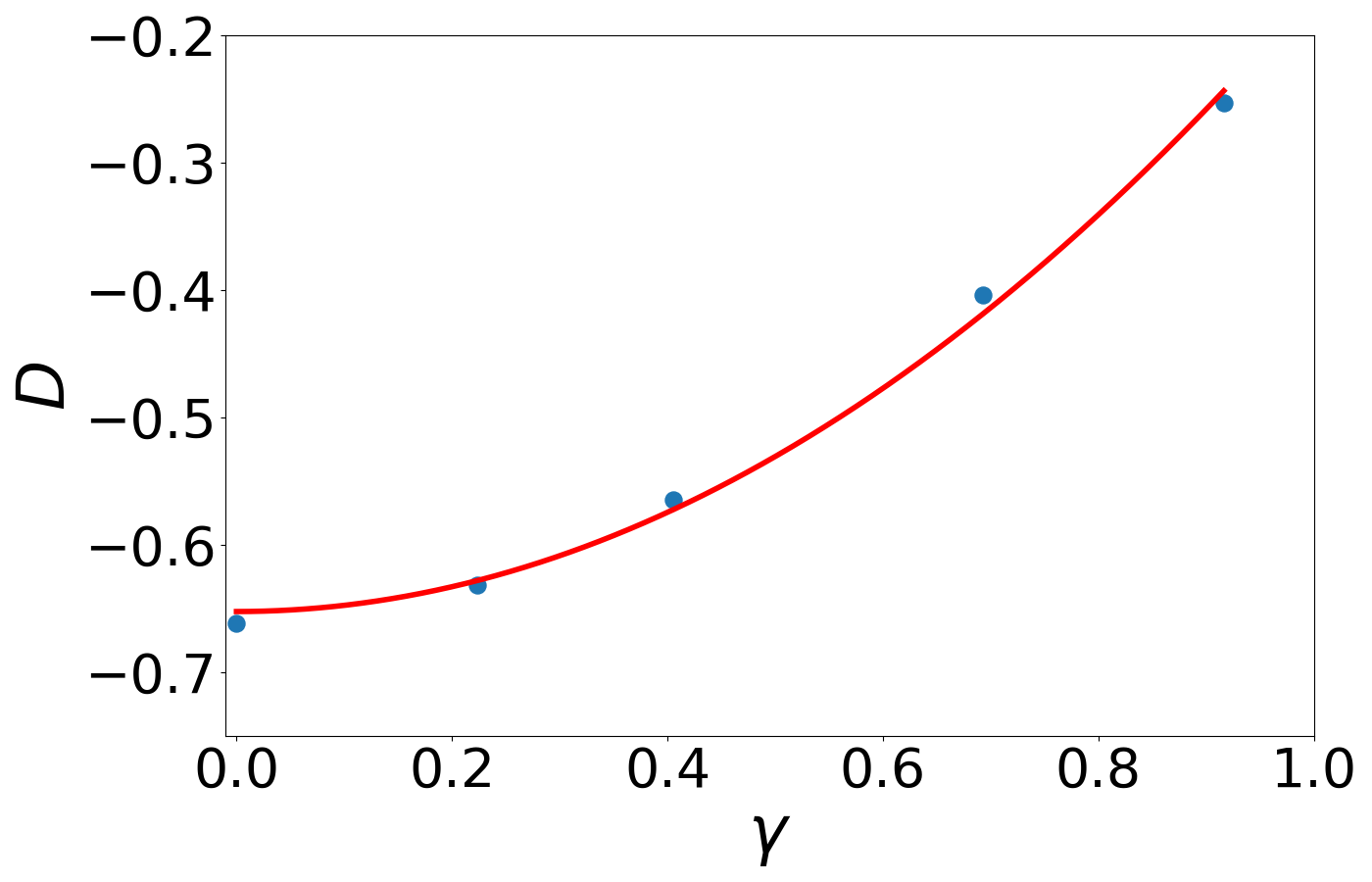"}
	
	(c)
	
	\caption{Anisotropy of the quantum Hall state vs. $\alpha$ for $C_4$ interaction anisotropy at $\nu=1/3$ in the presence of Coulomb interactions using (a) $q\rightarrow 0$ behavior and (b) maxima of the static structure factor. The red curves are best fits to Eq. \eqref{eq:fit_sigma_gamma}. We find (a) $s_1 = -0.357 \pm 0.007,\ s_3 = 0.116 \pm 0.010$ and (b) $\tilde s_1 = 0.453 \pm 0.001, \ \tilde s_3 = -0.076 \pm 0.009 $. Sub-figure (c) shows the dependence of $D$ on $\gamma$ as defined in Eq. \eqref{eq:D_definition}. The red curve is best fit to Eq. \eqref{eq:fit_D_gamma} with $d_0 = -0.652 \pm 0.008,\ d_2 = 0.487 \pm 0.018$.}
	\label{fig:ani_QH_Coulomb}
\end{figure}

We present the case of Coulomb interactions more closely. Fig. \ref{fig:ani_QH_Coulomb} shows the anisotropy in the $\nu=1/3$ state $\alpha_{\rm QH}$ vs. the interaction anisotropy $\alpha$. Using the definitions of Refs. \onlinecite{Krishna2019, Ippoliti2017b}:
\begin{gather}
\gamma \equiv \log \alpha\\
\sigma \equiv \log\alpha_{\rm QH}\\
D \equiv \frac{1}{2} \log \left[\frac{\lambda(\alpha)+\lambda(1/\alpha)}{2}\right] \label{eq:D_definition},
\end{gather}
we fit our data to the following forms:
\begin{gather}
	\sigma(\gamma) = s_1 \gamma + s_3 \gamma^3 \label{eq:fit_sigma_gamma}\\
	D(\gamma) = d_0 + d_2 \gamma^2 \label{eq:fit_D_gamma}
\end{gather}
From the small-$q$  behavior, we obtain:
\begin{gather}
	\sigma(\gamma) = -(0.357 \pm 0.007)\gamma + (0.116 \pm 0.010) \gamma^3\\
	D(\gamma) = -(0.652 \pm 0.008) + (0.487 \pm 0.018) \gamma^2
\end{gather}
The coefficients of $\gamma$ in $\sigma(\gamma)$  is significantly bigger in magnitude than $s_1 = 0.11$ found in Ref. \onlinecite{Krishna2019} for $C_4$-symmetric band mass anisotropy. This shows that unlike the case of twofold anisotropy where the band and interaction anisotropies can be mapped onto each other, the anisotropy in the interaction is a stronger effect than the band anisotropy for fourfold rotational symmetry. Based on this, one may also expect a similar behavior for higher-order anisotropies. Notice that the signs of $s_1$ in our work and Ref. \onlinecite{Krishna2019} are opposite. This is because the band mass anisotropy $\alpha_{\rm BM}$ defined in Ref. \onlinecite{Krishna2019} would lead to an \textit{effective} interaction-anisotropy opposite in sign to the one in Eq. \eqref{eq:ani_prescription}.

From the maxima of the static structure factor at $q = q_{\rm max}$, we get:
\begin{gather}
	\tilde\sigma(\gamma) = (0.453 \pm 0.001)\gamma - (0.076 \pm 0.009) \gamma^3
\end{gather}
This is bigger in magnitude and opposite in sign compared to one calculated using the small-$q$ behavior of $S(\bm q)$. This suggests that the correlations of the Laughlin state are modified differently at long distances than at short distances.

\section{$C_4$-symmetric interaction anisotropy at $\nu=1/2$\label{sec:1_2}}
In this section, we look at the effect of broken rotational symmetry on the composite-Fermi liquid (CFL) state at $\nu=1/2$ when the equipotentials of the interaction are $C_4$-symmetric. A natural way to define the anisotropy in this case is via the shape of the CF Fermi surface $k_F(\phi)$.\cite{Ippoliti2017} We make an ansatz that it lies along the curve ${k_F^2}_{\rm ani}(k_F,\phi) = \text{constant}$ where ${k_F}_{\rm ani}$ is defined along the lines of Eq. \eqref{eq:ani_prescription}. Assuming that the area of the Fermi surface is fixed by Luttinger's theorem in the 2D limit, it is given by:
\begin{gather}
	k_F^2(\phi) = \frac{2}{\left(\alpha_{CF} + \dfrac{1}{\alpha_{CF}}\right) + \left(\alpha_{CF} - \dfrac{1}{\alpha_{CF}}\right)\cos 4\phi}\label{eq:CFL_ani_FS}
\end{gather}

\begin{figure}
	
	\centering
	\includegraphics[width=2.8in]{"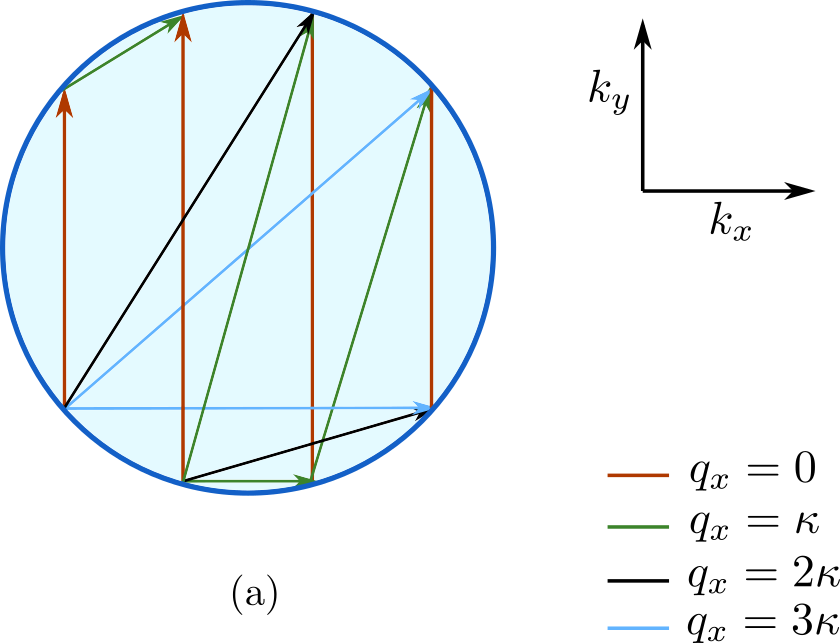"}

	\includegraphics[width=2.8in]{"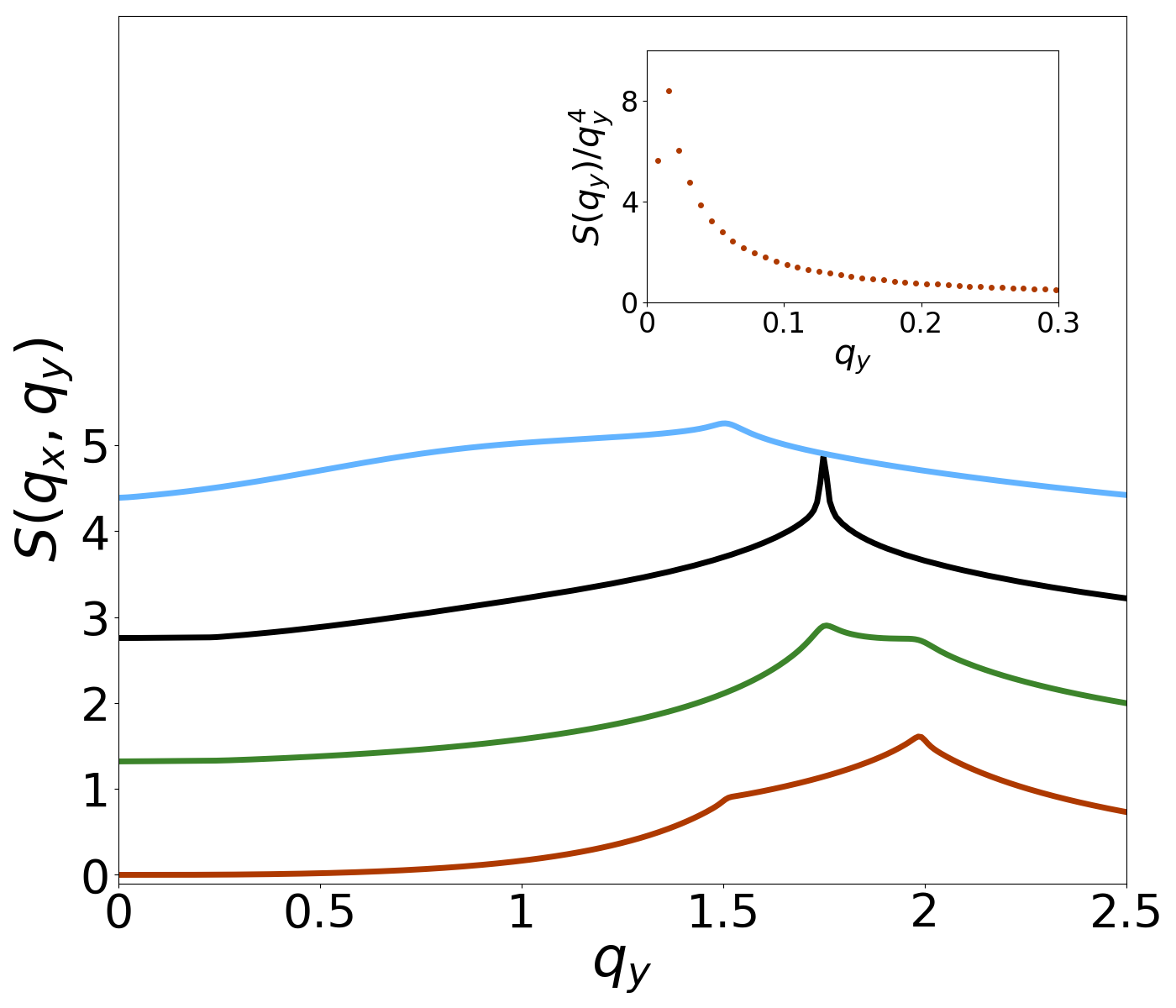"}
	
	(b)
	
	\caption{(a) Composite-fermion Fermi-surface with anti-periodic boundary conditions on a quasi-1D cylinder of circumference $L_x=12.5\ell_B$. The 2D-circular Fermi surface transforms to a set of four wires, shown in red, at $q_x = (n+1/2)\kappa$ where $\kappa = 2\pi/L_x$ and $-2 \leq n \leq 1$ is an integer. The set of unique particle-hole excitations across the Fermi-surface at $\bm q = \Delta \bm k$ are shown with arrows. (b) Guiding center static structure factor $S(\bm q)$ in the presence of $C_4$-symmetric anisotropic Coulomb interactions with $\alpha=0.67$. The singularities in $S(\bm q)$ are visible at $\bm q$-values described in (a). The inset shows $S(q_x=0, q_y)/q_y^4$ vs. $q_y$. Unlike the gapped QH states, it diverges in the limit $q_y \rightarrow 0$.}
	\label{fig:CFL_wires}
\end{figure}

Our DMRG simulations are performed in a quasi-1D geometry where the 2D Fermi sea transforms to a set of wires at discrete $k_x$ values uniformly spaced by $\Delta k_x = 2\pi/L_x$ (see Fig. \ref{fig:CFL_wires}), where $L_x$ is the circumference of the cylinder. The static structure factor contains singularities that correspond to excitations of particle-hole pairs via scattering across this discretized Fermi surface. Using methods described in Ref. \onlinecite{Geraedtsetal2015, Ippoliti2017, Ippoliti2017b}, we can sample points on the Fermi surface using these singularities and thus determine the anisotropy using Eq. \eqref{eq:CFL_ani_FS}. However, there is one caveat.
The quasi-1D geometry produces a systematic finite-size effect by modifying the 2D Luttinger's theorem. Luttinger's theorem on an infinite cylinder states that the sum of the lengths of the CFL Fermi-wires is fixed by the density of composite-fermions at $\nu=1/2$, i.e.,
\begin{gather}
	\sum_{i=1}^{N_{\rm wires}} Q_{y,i} = \frac{L_x}{2\ell_B^2} \label{eq:Luttinger_wire}
\end{gather}
where $Q_{y,i}$ is the length of the $i^{\rm th}$ wire. Since the 2D rotational symmetry of the $xy$-plane is broken down to a twofold discrete rotational symmetry, the set of CFL wires do not necessarily lie on a circle even for isotropic interactions. Therefore, only the relative lengths of the wires matters. 

\begin{figure}
	\centering
	\includegraphics[width=2.8in]{"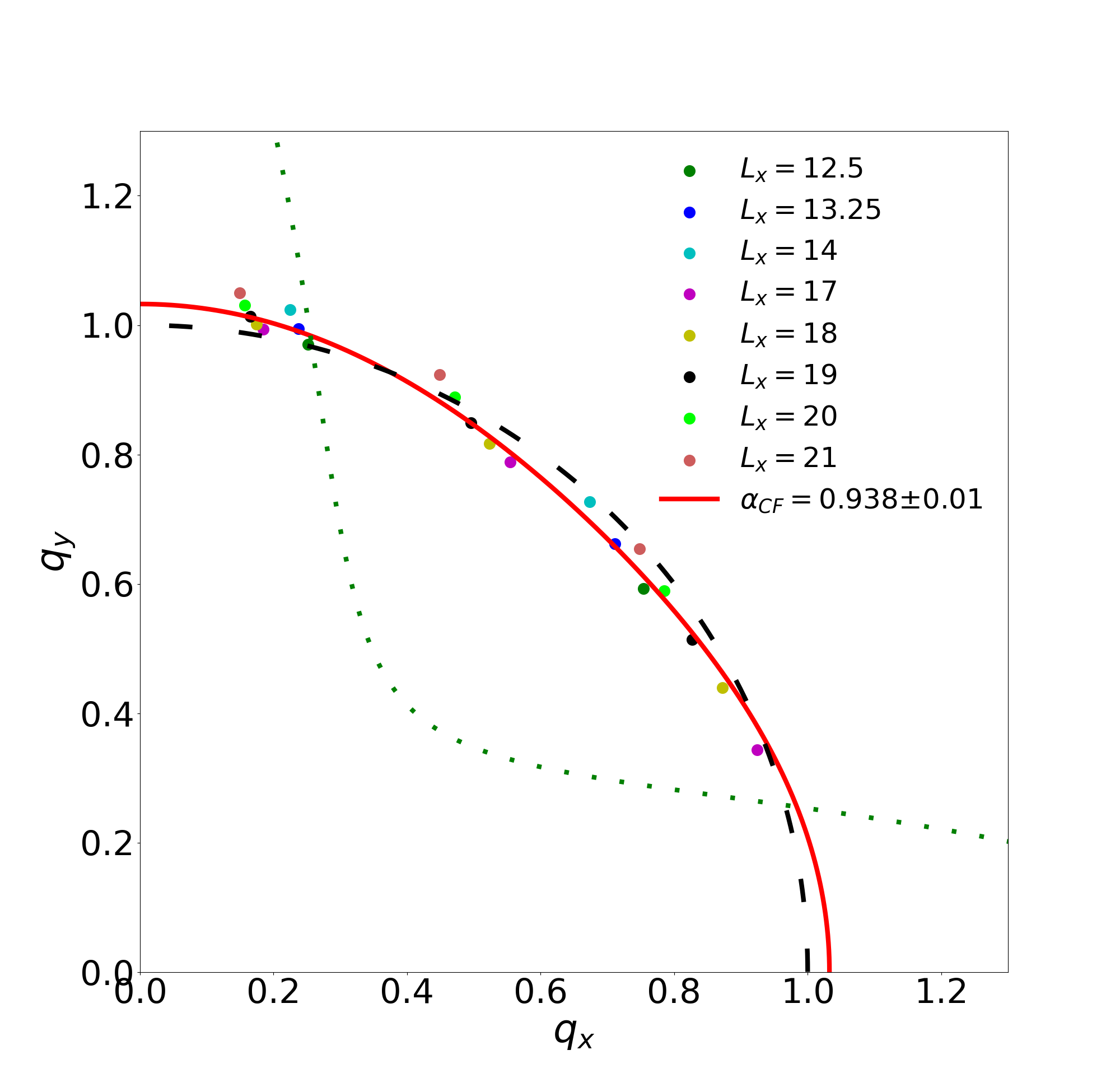"}
	
	(a)
	
	\includegraphics[width=2.8in]{"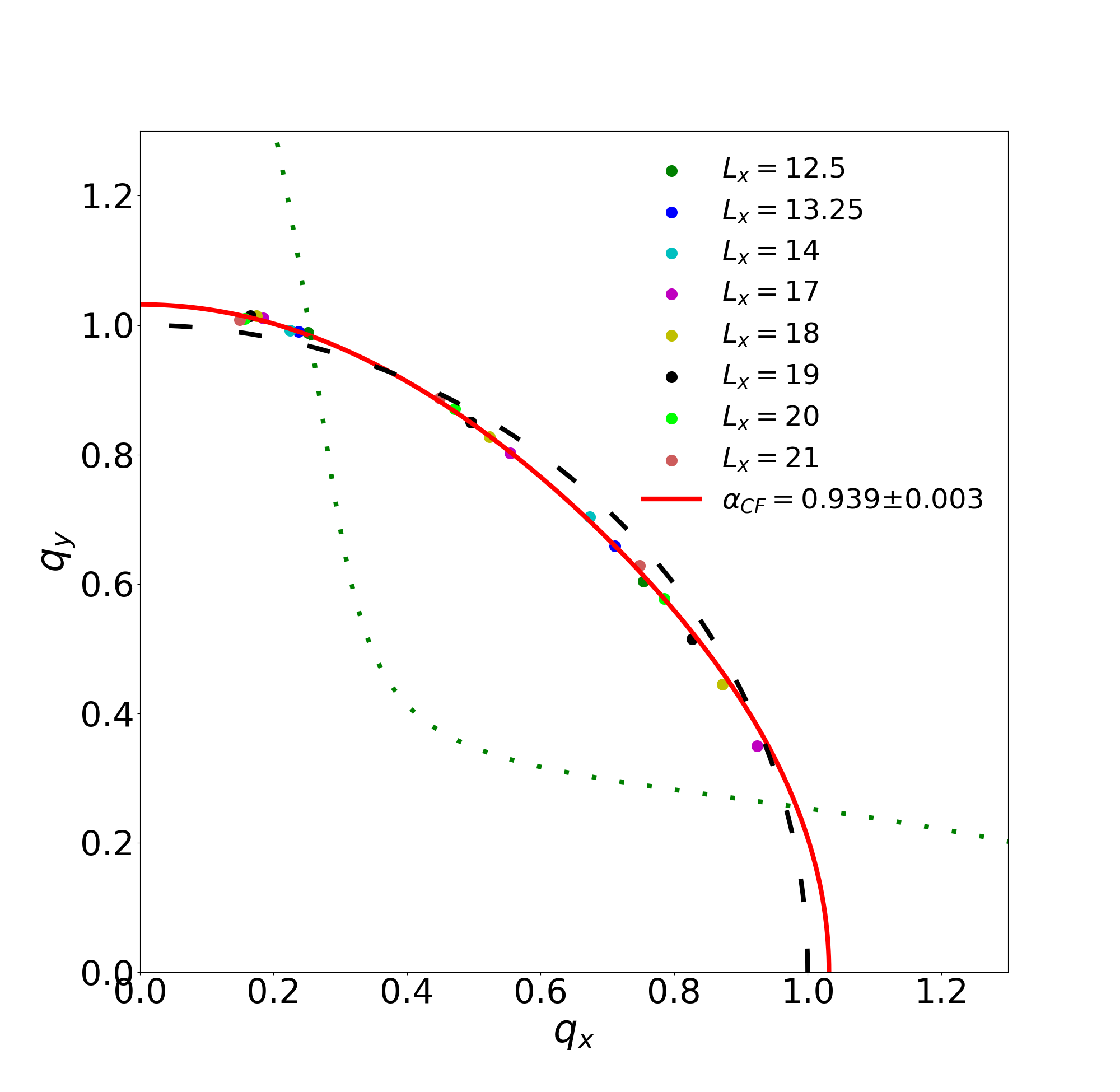"}
	
	(b)
	
	\caption{The effect of the scaling procedure described below Eq. \eqref{eq:Luttinger_wire} on the Fermi-surface determined using the singularities in the static structure factor at $\nu=1/2$. Applying this procedure on (a) reduces the error in $\alpha_{CF}$ by a factor of 3 as shown in (b). The interaction corresponds to $\eta=2$ in Eq. \eqref{eq:pow_law_interaction} and $\alpha=0.33$. Further, the dotted green curve shows the equipotential of the anisotropic interaction in the real-space.}
	\label{fig:unscaled_vs_rescaled_FS}
\end{figure}

To eliminate this elliptical non-area preserving deformation coming from the finite-size geometry, we modify our ansatz \eqref{eq:CFL_ani_FS}. We re-express the RHS in terms of $k_x$ and $k_y$ and scale $k_y$ by a to-be-determined scaling factor $b(L_x)$, i.e. $k_y \rightarrow b(L_x)\times k_y$. This can significantly reduce error in the measured anisotropies for some cases. One such example is shown in Fig. \ref{fig:unscaled_vs_rescaled_FS} where the error is reduced by a factor of 3. Notice that $\lim_{L_x\rightarrow \infty}b(L_x) =  1$.

\begin{figure}
	\centering
	\includegraphics[width=2.5in]{"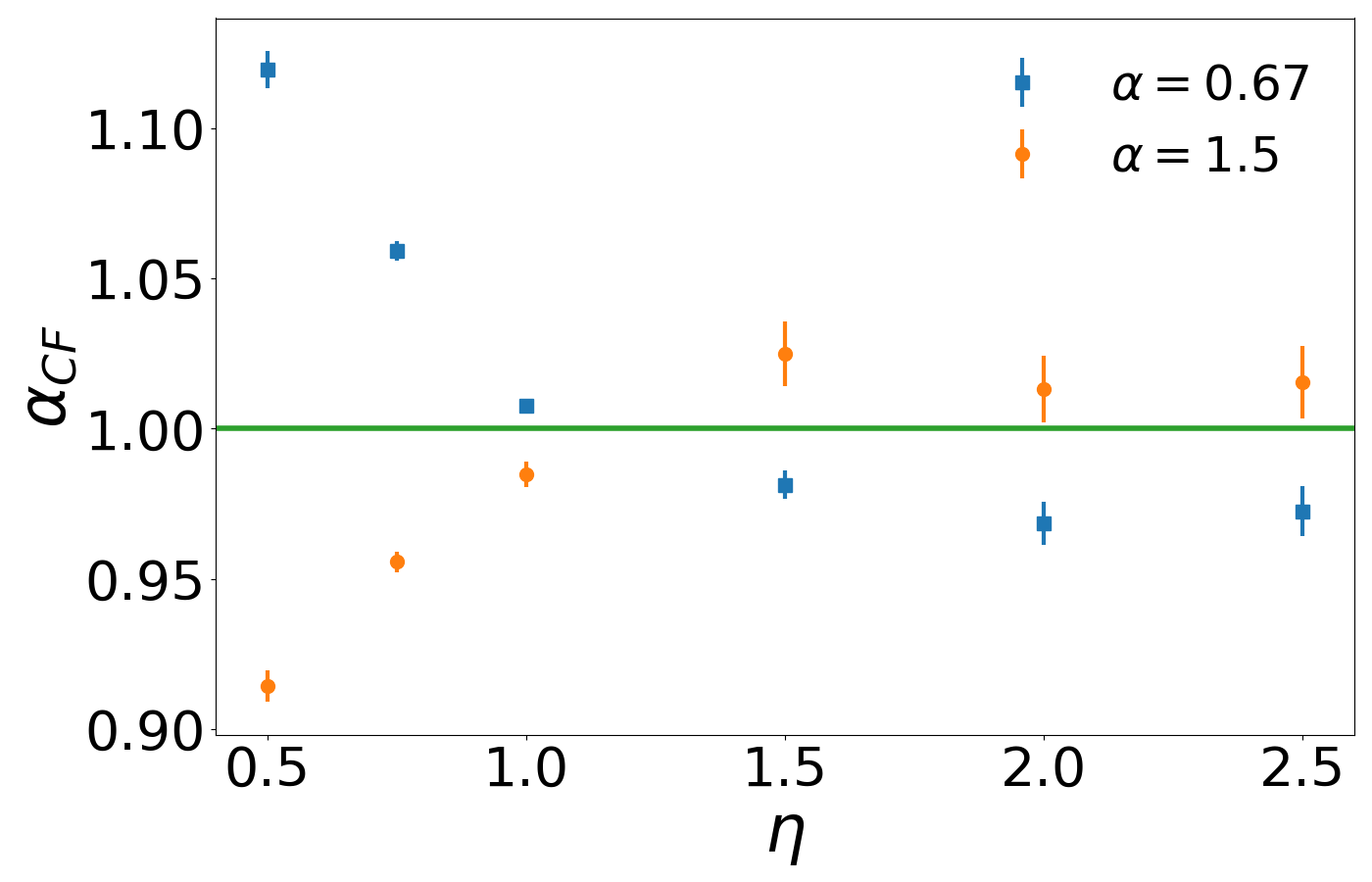"}

	\caption{The dependence of the anisotropy in the CF Fermi-surface $\alpha_{\rm CF}$ vs. the exponent of power-law interactions defined in Eq. \eqref{eq:pow_law_interaction} at $\nu=1/2$. The deformation in the Fermi surface changes sign at $\eta \approx 1.1$.}
	\label{fig:ani_CFL}
\end{figure}

Having described the methods, we now study the amount of anisotropy in the CFL Fermi surface vs. the exponent $\eta$ of the power law interactions. The results are plotted in Fig. \ref{fig:ani_CFL}. Quite surprisingly, we observe that the deformation in the Fermi-surface changes sign at $\eta \approx 1.1$.

\begin{figure}
	\centering

	\includegraphics[width=2.5in]{"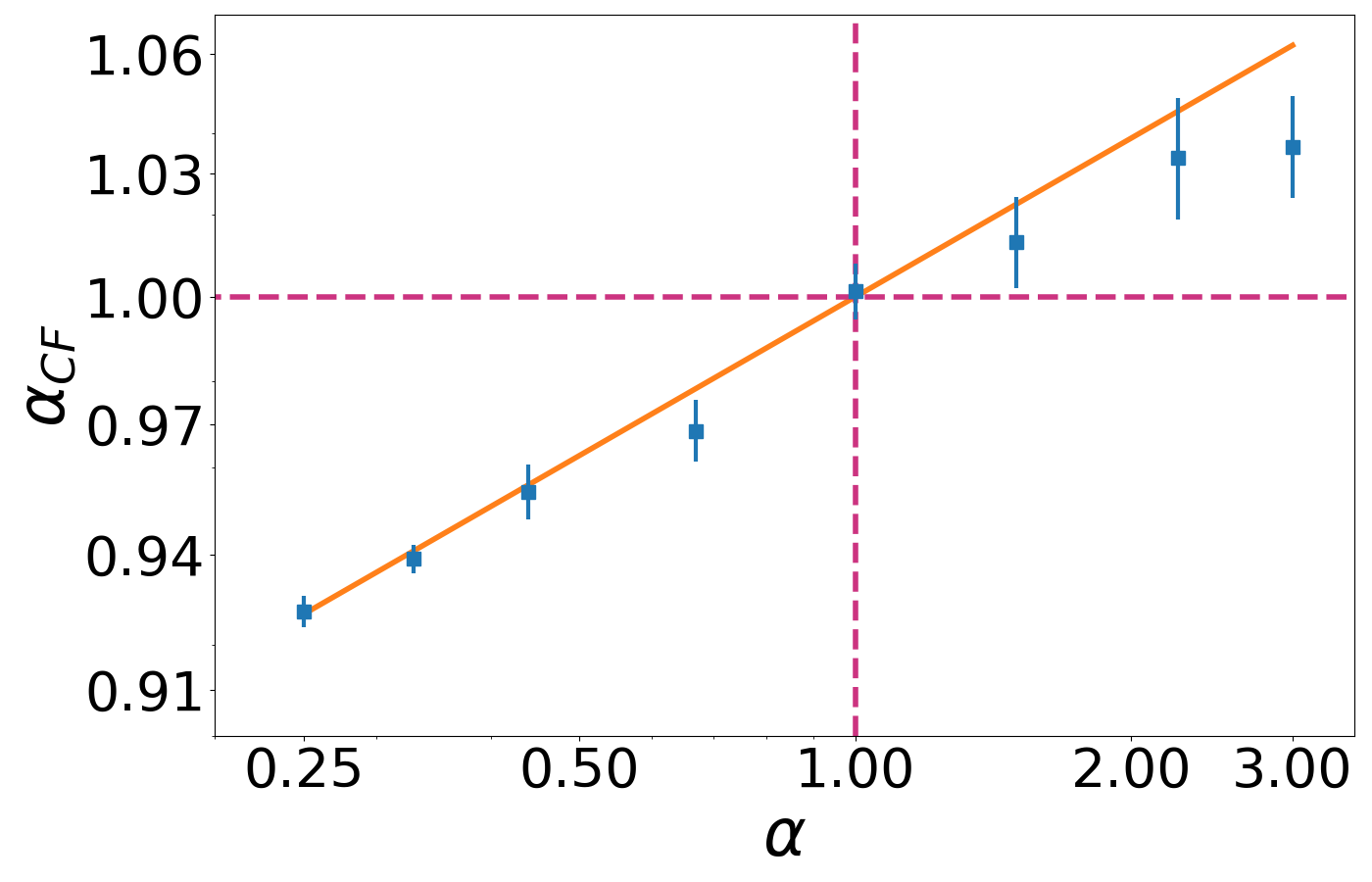"}
	
	(a)
	
	\includegraphics[width=2.5in]{"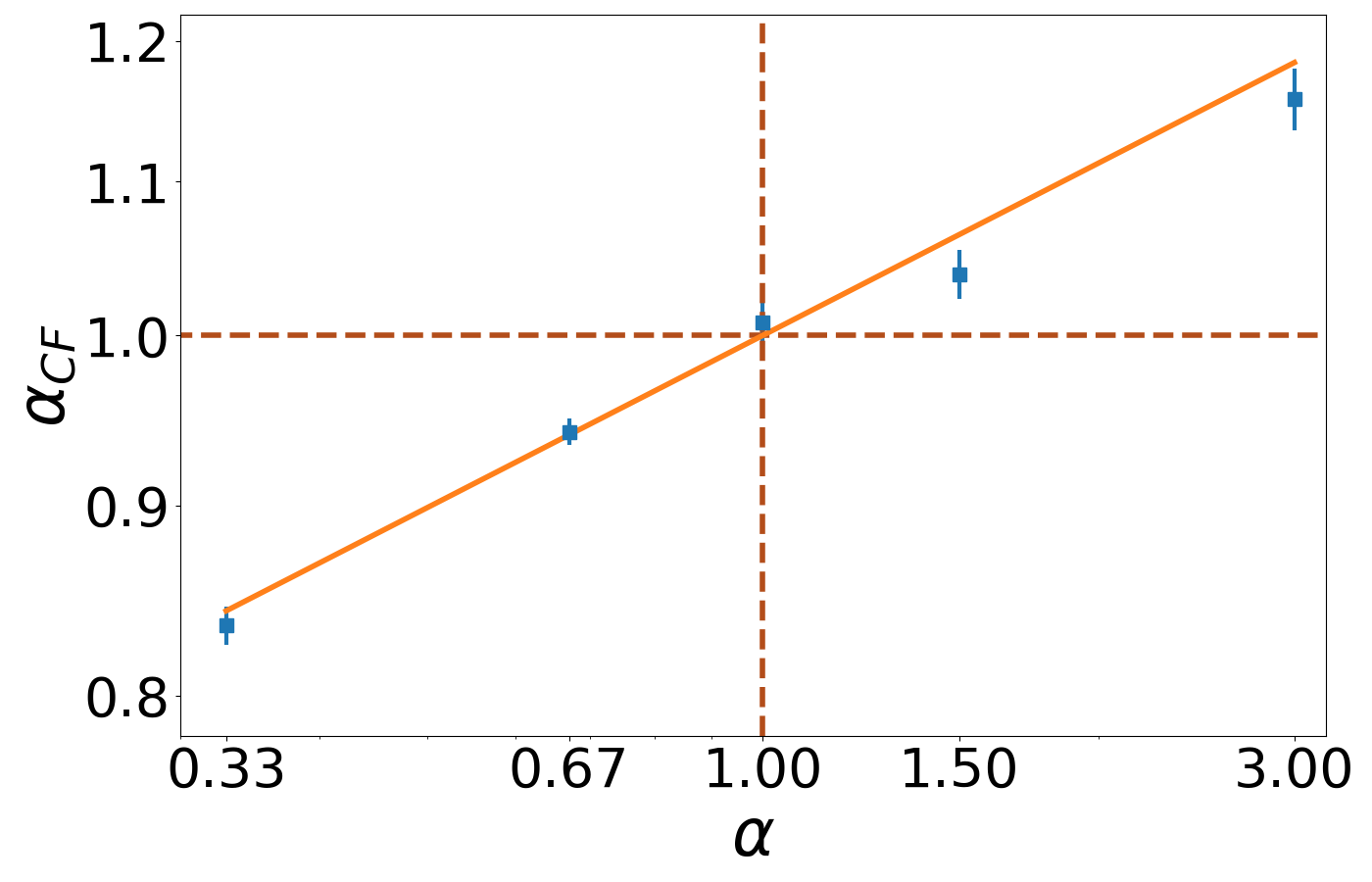"}
	
	(b)
	
	\caption{$\alpha_{CF}$ vs $\alpha$ for short-range interactions with (a) $V(r) = 1/r^2$, (b) $V(r) = e^{-r^2/2\ell_B^2}$ for stronger anisotropies than used in Fig. \ref{fig:ani_CFL}. The error in $\alpha_{CF}$ is small compared to the magnitude of the deformation thus confirming the sign-reversal observed in Fig. \ref{fig:ani_CFL}. The orange line shows the best fit to $\alpha_{\rm cf} = \alpha^\beta$ with (a) $\beta = 0.055 \pm 0.002$, (b) $\beta = 0.154 \pm 0.008$.}
	\label{fig:ani_CFL_short_range}
\end{figure}

The deformation parameter $\sigma \equiv \log\alpha_{\rm CF}$ is significant for $\eta < 1.1$. However, it is quite small for $\eta >1.1$. As such, it is not clear if the sign-reversal is a real effect or an artifact of numerics coming from finite sizes, slow convergence etc. To provide the observation of sign-reversal a firmer ground, we analyze the $\nu=1/2$ state at short-range interactions for stronger anisotropies. In Fig. \ref{fig:ani_CFL_short_range}, we plot $\alpha_{\rm CF}$ vs. $\alpha$ for the power-law interaction corresponding to $\eta=2$ and the gaussian interaction: $V(r) = e^{-r^2/2\ell_B^2}$. The deformation in the Fermi-surface is found to become stronger and exceed the error bars comfortably for large interaction-anisotropies. This confirms that the deformation in the CFL Fermi-surface has opposite signs for long-range and short-range interactions.

This finding is in contrast with the results obtained for $\nu=1/3$ (Fig. \ref{fig:ani_QH}(b)) in the previous section where the deformation is always in the same orientation. Similarly, no such behavior was encountered for the case of $C_2$-symmetric anisotropy.\cite{Ippoliti2017, Krishna2019} Further, in Ref. \onlinecite{Ippoliti2017b}, it was found that the CF Fermi surface has a very weak deformation when a $C_4$-symmetric band mass anisotropy is introduced in the presence Coulomb interactions. This may be related to the fact that Coulomb interactions are close to the point where the sign-reversal is observed. More importantly, our result suggests that the short-distance and long-distance parts of the electron-electron interaction have opposite effects when the interaction has a $C_4$-symmetric anisotropy at $\nu=1/2$.

From a more theoretical standpoint, we can ask why the deformation in the shape of the Fermi-surface changes sign. To answer this question, one needs a theory that can be used to determine the shape of the CF Fermi-surface for anisotropic interactions. We can draw some intuition from the $\nu_B=1$ bosonic state which is another realization of the CFL state where the composite-fermions are formed by attaching one-flux quantum to each boson. At this filling fraction, Pasquier-Haldane construction,\cite{PasquierHaldane, Read1998, Dong2020} inspired by the dipole-interpretation of the CFL state, provides a mechanism for the formation of a compact CF Fermi-surface. As explained in Refs. \onlinecite{PasquierHaldane, Read1998, Dong2020}, it involves three effects: the energy of the dipole, the dipole-dipole interaction and certain constraints on an enlarged Hilbert space. In Ref. \onlinecite{Kumarunpub}, the first two terms are analyzed and they are found to compete against each other. However, the second term wins over the first for all values of $\eta$ and the sign of the distortion agrees with the $\eta < 1.1$ side of the results shown in Fig. \ref{fig:ani_CFL}. Nevertheless, for a more faithful understanding of the results of this section, a generalized version of Pasquier-Haldane construction, that can work at $\nu=1/2$ of electrons, needs to be developed.

\section{Discussion\label{sec:discussion}}
In summary, we calculated the anisotropy in the quantum Hall states at $\nu=1/3$ and $\nu=1/2$ when the interaction breaks the continuous rotation symmetry to a fourfold discrete rotational symmetry. We found that generally the anisotropy in the quantum Hall (QH) states characterized by $\alpha_{\rm QH}$ is much weaker than the anisotropy in the interaction characterized by $\alpha$. Also, it becomes weaker as the interactions are made short-ranged. 

At $\nu=1/3$ in the presence of $1/r$ interactions, the quantum Hall state experiences a bigger deformation for anisotropic interactions with fourfold symmetry than a corresponding anisotropy of the band. This is relevant to experiments where both effects are present. However, it is unclear if and how the anisotropy can be measured. The angle resolved magnetoroton dispersion may provide one such possibility.

An interesting finding is that the deformation in the composite-fermion (CF) Fermi-surface at $\nu=1/2$ changes sign as one changes the range of electron-electron interactions. This suggests that the long-distance and short-distance parts of the interaction have opposite effects in the presence of the $C_4$-symmetric anisotropy. Such a behavior is not observed either in the $\nu=1/3$ state or the $C_2$-symmetric anisotropy at $\nu=1/2$.\cite{Ippoliti2017, Krishna2019} This might be relevant in experimental systems with square symmetry, such as GaAs quantum wells with the 2DEG plane oriented perpendicular to one of the principal axes of the crystal. However, this would require the magnetic length to be of the same order as the lattice spacing since the dielectric tensor becomes isotropic at long distances.


On the theoretical front, generalized versions of anisotropic model wavefunctions developed in Refs. \onlinecite{Haldane2011, Qiu2012, Yang2012, Balram2016, Ciftja2017} can provide a complimentary perspective to our results. Further, developing a generalized version of the Pasquier-Haldane construction appears to be a promising direction to study the interaction dependent anisotropy at $\nu=1/2$.

\section{Acknowledgment}
We thank Matteo Ippoliti and Akshay Krishna for discussions. The iDMRG numerical computations were carried out using libraries developed by Roger Mong, Michael Zaletel and the TenPy collaboration. This work was supported by DOE BES Grant No. DE-SC0002140.


\end{document}